\documentstyle[preprint,aps,epsf,floats,rotating]{revtex}

\newcommand{\beq}{\begin{equation}}
\newcommand{\eeq}{\end{equation}}
\newcommand{\bea}{\begin{eqnarray}}
\newcommand{\eea}{\end{eqnarray}}
\newcommand{\ben}{\begin{eqnarray*}}
\newcommand{\een}{\end{eqnarray*}}

\def\D0{D\O}

\newcommand{\epsl}{\not\!\epsilon}
\newcommand{\psl}{\not\!p}
\newcommand{\ksl}{\not\!k}
\newcommand{\lsl}{\not l}
\newcommand{\vsl}{\not\!v}
\newcommand{\zsl}{\not\!z}

\tightenlines

\begin{document}

\title{$B$ Production Asymmetries in Perturbative QCD}

\author{Eric Braaten\footnote{braaten@mps.ohio-state.edu},
Yu Jia\footnote{yjia@mps.ohio-state.edu}
and Thomas Mehen\footnote{mehen@mps.ohio-state.edu}
\footnote{address after Jan. 1, 2002:
Department of Physics, Duke University, Durham NC 27708.} }

\address{Department of Physics, The Ohio State University, Columbus, OH
43210}

\date{\today}

\maketitle

\begin{abstract}

This paper explores a new mechanism for $B$ production in which a $\overline{b}$ quark combines with a
light parton from the hard-scattering process before hadronizing into the $B$ hadron. This
recombination mechanism can be calculated within perturbative QCD up to a few nonperturbative
constants. Though suppressed at large transverse momentum by a factor $\Lambda_{\rm QCD} m_b/p_\perp^2$
relative to $\overline{b}$ fragmentation production, it can be important at large rapidities. A
signature for this heavy-quark recombination mechanism in $p\overline{p}$ colliders is the presence of
rapidity asymmetries in $B$ cross sections. Given reasonable assumptions about the size of
nonperturbative parameters entering the calculation, we find that the asymmetries are only significant
for rapidities larger than those currently probed by collider experiments.

\end{abstract}

\bigskip

\thispagestyle{empty}

\newpage

\section{Heavy quark production}

Perturbative QCD calculations of heavy hadron production are based on the factorization theorems
of QCD \cite{Collins:1986gm}. These theorems imply that, for sufficiently large transverse
momentum, production of a heavy-light hadron occurs in two stages: first, a heavy quark is
produced in a short-distance partonic process; second, the heavy quark fragments into the hadron.
They allow cross sections to be factorized into partonic cross sections and fragmentation
functions. For example, the formula for the $B$ production cross section in a gluon-gluon
collision is
\bea
\label{fact}
d \sigma[gg\rightarrow B+X]= d \hat{\sigma}[gg \rightarrow b\overline{b}] \otimes
D_{\overline{b}\rightarrow B}  \, .
\eea
(We denote hadrons with $b$ quarks as $\overline{B}$ and hadrons with $\overline{b}$ antiquarks as
$B$.) The symbol $\otimes$ represents convolution over the final-state momentum fraction, $z$, of the
$B$ hadron in the $\overline{b}$ quark jet. The fragmentation function $D_{\overline{b}\rightarrow
B}(z)$ is a universal function that involves long-distance physics and hence cannot be computed
perturbatively. On the other hand, the partonic cross section  $d \hat{\sigma}[gg\rightarrow
b\overline{b}]$ depends only on short-distance physics and can be  calculated in perturbation theory.
To obtain cross sections for hadronic collisions, partonic cross sections such as Eq.~(\ref{fact}) are
convolved with parton distribution functions for the initial-state partons.

The QCD factorization theorems are exact only in the limit of large transverse momentum, $p_\perp
\rightarrow \infty$. For finite $p_\perp$, one expects corrections that are suppressed by powers
of $\Lambda_{\rm QCD}/p_\perp$.  In this paper we will argue that there are $O(\Lambda_{\rm QCD}
m_b/p_\perp^2)$ corrections that are calculable within perturbative QCD up to a few multiplicative
constants. These corrections come from processes in which a light quark, $q$, participating in the
hard-scattering process combines with the $\overline{b}$ and they subsequently hadronize into a
state including the $B$ hadron. We will refer to this as a $\overline{b}q$ recombination process.
In $p\overline{p}$ colliders, this process can be distinguished qualitatively from $\overline{b}$
fragmentation because it gives interesting rapidity asymmetries in $B$ hadron production. Though
not important in the central rapidity region, $\overline{b}q$  recombination can actually give a
larger contribution than $\overline{b}$ fragmentation at large rapidities.

Before describing this mechanism in detail, it is worthwhile to briefly summarize the current
status of $B$ production. Theoretical calculations of heavy quark production have been performed
to NLO in perturbative QCD \cite{Nason:1989zy,Beenakker:1989bq,Beenakker:1991ma}. Comparison of
these perturbative QCD calculations with collider data is reviewed in
\cite{Frixione:1997ma,Mangano:1997ri}. At the Tevatron
\cite{Abe:1993sj,Abe:1993vw,Abachi:1995kj,Abe:2000ac,Abbott:2000wu}, the shape of the $p_\perp$
distribution is in agreement with experiments, while its normalization appears to be somewhat
underpredicted by the theory. However, even at NLO the theoretical prediction suffers from
considerable uncertainty due to dependence on the renormalization scale. Setting the
renormalization scale, $\mu_R$, equal to $(p_\perp^2+m_b^2)^{1/2}$ gives a prediction that is
roughly a factor of 2 below experiment, while for $\mu_R \approx (p_\perp^2+m_b^2)^{1/2}/2$  the
theoretical prediction can accomodate the data. The strong sensitivity to the renormalization
scale indicates that  higher order perturbative corrections are important. Recent studies that
attempt to resum  logarithmically enhanced higher order corrections can be found in
\cite{Cacciari:1994mq,Olness:1999yc,Ball:2001pq,Field:2002da}. The resummation of higher order
terms of the form  $\alpha_s^n\, {\rm ln}^n(p_\perp/m_b)$ and $\alpha_s^n \,{\rm
ln}^{n+1}(p_\perp/m_b)$ gives results with reduced scale dependence and the prediction of the
resummed calculation is similar to that of the fixed NLO calculation with the lower values of
$\mu_R$. Thus it seems at least possible that higher order perturbative QCD calculations would
account for the shape and normalization of the experimentally observed $p_\perp$ distribution.

Most measurements of heavy quark production focus on the central rapidity region.  Less is known about
the cross section in the forward regions. The $\D0$ collaboration \cite{Abachi:1996uy} has measured the
production of muons coming from $b$ decay in the rapidity range $2.4< |y^\mu| <3.2$. They found that
the excess over the theoretical prediction in the forward region is greater than in the central region,
indicating perhaps that large rapidity $b$  production is not as well understood. If one chooses
$\mu_R$ so that the central region is described correctly, the NLO calculation of the cross section at
large rapidity is too small by a factor of 2. It has been suggested that modification of the heavy
quark fragmentation function \cite{Mangano:1997ri} or a reorganization of the perturbative calculation
\cite{Olness:1999yc} could modify the forward-to-central rapidity ratio. However, CDF \cite{Abe:2000ac}
measurements of $b\overline{b}$ rapidity correlations agree well with NLO QCD calculations. CDF
measured the ratio of central ($|y| <0.6$) to forward $(2.0 < |y| <2.6)$ $b$ production subject to the
constraint that the other $b$ was produced with $|y|<1.5$ and that both $b$'s had $p_\perp > 25 \,{\rm
GeV}$. It is important to point out that one cannot directly compare the $\D0$ and CDF experiments
because a) they involve  slightly different rapidity regions,  b) $\D0$ makes no cuts on the second $b$
produced in the event and c) most importantly, the $\D0$ experiment uses a much smaller $p_\perp$ cut
of 5 GeV.

\section{Heavy-quark recombination mechanism}

Our investigation of novel mechanisms for $B$ production is motivated in part by the possibility that
there are important contributions to $B$ production at large rapidity that are not accounted for in
Eq.~(\ref{fact}).  This paper will consider contributions to $B$ production in which a light quark
involved in the short-distance process combines with the $\overline{b}$ and they subsequently hadronize
into the final-state $B$ meson. This $\overline{b}q$ recombination contribution is distinct from what
is included in standard fragmentation calculations, where the parton process $qg\rightarrow
b+\overline{b}+q$ is taken into account at next-to-leading order.  Over most of the phase space, the
$q$ in the final state produces a jet that is distinct from those produced by the $b$ and
$\overline{b}$. There is a small contribution from the corner of phase space where the $q$ has small
momentum in the rest frame of the $\overline{b}$. However, in this region, the $q$ and $\overline{b}$
can bind and large nonperturbative effects can enhance the cross section. Our mechanism takes these
nonperturbative effects into account.

The word ``recombination'' in high energy physics dates back to the {\it parton recombination model}
invented by Das and Hwa in 1977 to describe the production of mesons at low $p_\perp$ in hadronic
collisions \cite{Das:1977cp}. In this model, most of the momentum of the meson comes from the momentum
of a valence parton in one of the colliding hadrons. The meson is formed by that valence parton
recombining with a sea parton from that same hadron. Our recombination mechanism for heavy meson
production is analogous in that it also allows a valence parton to deliver most of its momentum to the
heavy meson, although in this case the valence parton combines with a heavy quark created in the hard
scattering. Our recombination mechanism is also analogous to a ``higher twist'' mechanism for the
inclusive production of light mesons at large $p_\perp$ studied by Berger, Gottschalk, and Sivers
\cite{Berger:1981qg}. This mechanism also involves a parton from the colliding hadron that is involved
in the hard scattering and ends up as a constituent of the large-$p_\perp$ meson.  The cross section
is suppressed by $\Lambda_{\rm QCD}^2/p_\perp^2$ relative to the fragmentation mechanism, and the
nonperturbative factor involves a function that describes the distribution of the light-cone momentum
in the meson. In our mechanism for heavy meson production, the cross section is suppressed only by
$\Lambda_{\rm QCD} m_b/p_\perp^2$, and the nonperturbative factor can be reduced to a few numbers. For
$p_\perp \sim m_b$, the contribution from our mechanism is suppressed only by $\Lambda_{\rm
QCD}/m_b$.  In this region of $p_\perp$, there are other contributions suppressed by $\Lambda_{\rm
QCD}/m_b$ in which the parton that combines with the heavy quark is a spectator not involved in the
hard scattering \cite{Brodsky:1987cv}. String fragmentation models for hadronization provide an
explicit model for these contributions \cite{Norrbin-Sjostrand}. Another mechanism that gives
contributions at $p_\perp \gg m_b$ that are suppressed by $\Lambda_{\rm QCD}^2/m_b^2$ is intrinsic
heavy flavor \cite{Vogt:1992ki,Vogt:1994zf}. These other mechanisms may give contributions to the
integrated cross section that are comparable to that of our heavy-quark recombination mechanism, but
they should be smaller for $p_\perp \gg m_b$. Even for $p_\perp \sim m_b$, our mechanism may be
numerically important in certain kinematic regions, such as large rapidity, because it allows most of
the momentum of a valence parton to be delivered to the heavy meson.

To motivate our recombination mechanism for heavy mesons, it is helpful to make  an analogy with
quarkonium production \cite{Bodwin:1995jh,Braaten:1996pv,Beneke:1997av,Kramer:2001hh}. Quarkonium
production cross sections also obey a factorization formula:
\bea\label{onia}
d \sigma[ij\rightarrow H+X] = \sum_n d\hat{\sigma}[ij \rightarrow (Q\overline{Q})^n+X]
\langle 0| {\cal O}_n^H |0\rangle \, ,
\eea
where $i,j$ represent initial-state partons and $Q\overline{Q}$ represents the heavy
quark-antiquark pair. The $Q\overline{Q}$ are produced  in the short-distance process with small
relative momentum  in a state with definite angular momentum and color quantum numbers, which are
collectively denoted by $n$. The $Q\overline{Q}$ pair in the state $n$ subsequently evolves into a
final hadronic state that includes the quarkonium, $H$. The probability for this is given by a
vacuum matrix element of a local operator in NRQCD, denoted by ${\cal O}_n^H$. If the
$Q\overline{Q}$ produced in the short-distance process is a color-singlet and has the same angular
momentum quantum numbers as the quarkonium being produced, the matrix element has a physical
interpretation in terms of the nonrelativistic wave function of the $Q\overline{Q}$ pair in the
bound state.

The quarkonium factorization formula suggests that there are
analogous  contributions to $B$ production:
\bea\label{bjm}
d\sigma[q g \rightarrow B+X ] &=& \sum_n d\hat{\sigma}[q g \rightarrow (\overline{b}q)^n+b]
\, \rho[(\overline{b}q)^n \rightarrow B]\, .
\eea
(A similar mechanism for $\overline{B}$ production is obtained by replacing $b \leftrightarrow
\overline{b}$ and $q \rightarrow \overline{q}$). The notation $(\overline{b}q)^n$ means that the
$q$ has small momentum in the $\overline{b}$ rest frame and that the $\overline{b}$ and the $q$
are in a state with definite color and angular momentum quantum numbers specified by $n$. The
factor $\rho[(\overline{b}q)^n \rightarrow B]$ represents the probability for $(\overline{b}q)^n$
to evolve into a final state that includes the $B$ hadron. We refer to the process in
Eq.~(\ref{bjm}) as $\overline{b}q$ recombination.

Note that the color and spin quantum numbers of the $(\overline{b} q)^n$ need not coincide with
those in the quark model for the $B$ hadron. The $(\overline{b} q)^n$ can evolve into a state that
contains the $B$ hadron and an arbitrary number of partons that have small momentum in the $B$
hadron rest frame. In this nonperturbative transition, soft gluons can  radiate color and flip the
spin of the light quark so the $(\overline{b}q)^n$ produced in the short-distance process need not
be a color-singlet and its angular momentum quantum numbers can differ from those of the
final-state $B$ hadron. In quarkonia production, these nonperturbative transitions are governed by
the multipole expansion and the nonperturbative matrix elements appearing in Eq.~(\ref{onia})
scale with definite powers of the typical velocity, $v$, of the heavy quarks in the bound state.
Those transitions for which the color and angular momentum quantum numbers of the
$(Q\overline{Q})^n$ do not match those of $H$ are suppressed by powers of $v$, and are therefore
controlled by the heavy quark mass. We expect no such suppression for the parameters
$\rho[(\overline{b}q)^n \rightarrow B]$ in  Eq.~(\ref{bjm}).

Furthermore, it is important to observe that the $B$ meson need not have the same light quark
flavor quantum numbers as the $(\overline{b} q)^n$ pair that is produced in the short-distance
process.  The hadronization process can involve the creation of a light quark-antiquark pair,
$q^\prime\overline{q}^\prime$, and the $\overline{b}$ can bind with the $q^\prime$ to form a $B$
meson with a different light quark flavor from the $(\overline{b} q)^n$ that is produced in the
hard scattering.  This is an important difference from quarkonium production, where the total
heavy quark and heavy antiquark numbers are separately conserved because the hadronization process
is governed by long-distance physics. Light quark-antiquark pair production could of course have a
nonperturbative suppression factor. For instance, there is such a suppression factor in the large
$N_c$ limit of QCD. Thus a $(\overline{b} u)^n$ could prefer to hadronize into a $B^+$ rather than
a $B^0$.  If such a suppression factor exists, it is not controlled by the heavy quark mass and
should therefore be roughly the same for $b$ and $c$ hadrons. The flavor dependence of the
nonperturbative factor  $\rho[(\overline{b}q)^n \rightarrow B]$ is important for predictions of
flavor-specific $B$ hadron production cross sections. In this paper, we will for simplicity
consider only cross sections that are summed over all $B$ hadrons. For this inclusive cross
section, the nonperturbative factor in Eq.~(\ref{bjm}) can therefore be denoted simply by
$\rho_n$.

It is only necessary to consider $(\overline{b}q)^n$ with relative orbital angular momentum $L=0$.
We will show below that the recombination  constants $\rho[(\bar b q)^n \to B]$ for $L=0$ are
proportional to $\Lambda_{\rm QCD}/m_b$, so the recombination contributions to the cross section
are suppressed by $\Lambda_{\rm QCD} m_b/p_\perp^2$. Producing $(\overline{b}q)^n$ in a higher
orbital angular momentum state requires keeping additional factors of the light quark momentum in
the hard-scattering matrix element. The light quark momentum in the bound state is $O(\Lambda_{\rm
QCD})$ and these additional factors of $\Lambda_{\rm QCD}$ must be compensated by powers of $m_b$
or $p_\perp$. So up to corrections suppressed by additional powers of $\Lambda_{\rm QCD}/m_b$ or
$\Lambda_{\rm QCD}/p_\perp$, it is sufficient to compute production of  $(\overline{b} q)^n$ in
the spectroscopic states $^1S_0^{(1)},\,{}^3S_1^{(1)},\,{}^1S_0^{(8)}$  and  $^3S_1^{(8)}$, where
the superscript in parantheses indicates either color-singlet or color-octet. Furthermore, heavy
quark spin symmetry \cite{Isgur:1989vq,Isgur:1990ed} relates the transition amplitudes of $^1S_0$
and $^3S_1$ states. Thus the calculation involves only two nonperturbative parameters, $\rho_1$
for color-singlet and $\rho_8$  for color-octet production. Heavy quark flavor symmetry can also
be used to relate the production parameters for $b$ hadrons to those for $c$ hadrons.

In most regions of phase space, $\overline{b}$ fragmentation dominates over the $\overline{b}q$
recombination mechanism. However, the $\overline{b}q$ recombination cross section is peaked in the
forward region where the  $(\overline{b} q)$ pair emerges in approximately the same direction as the
initial-state light quark. As shown below,
it is possible for $\overline{b}q$ recombination to be comparable to
$\overline{b}$ fragmentation in the forward region.  In addition to enhancing the cross section,
recombination leads to an interesting rapidity asymmetry in $B$ hadron production at $p\overline{p}$
colliders. Let the proton direction define positive rapidity. If the light quark in the initial state
is from the proton, the $(\overline{b} q)$, and therefore the $B$ hadron, will emerge preferentially
with large positive rapidity, while if the initial-state light quark comes from the antiproton, the $B$
hadron will prefer to have negative rapidity. Since the proton has more $u,d$ than
$\overline{u},\overline{d}$, there should be an excess of $(\overline{b} u)$ and $(\overline{b}d)$ at
positive rapidity, while the $\overline{u},\overline{d}$ in the antiproton should lead to an excess of
$(b\overline{u})$ and $(b\overline{d})$ at negative rapidity.  This rapidity asymmetry is a striking
prediction of this mechanism.

Rapidity asymmetries have also been predicted on the basis of string fragmentation models for
hadronization\cite{Norrbin-Sjostrand,Norrbin:2000jy}. In these models, the asymmetry arises from a
process called the ``beam drag effect'' in which the $\overline{b}$ produced by the hard scattering
combines with a spectator quark from an initial-state hadron. Since the beam drag effect necessarily
involves soft momentum scales, this nonperturbative contribution is part of the  corrections to
Eq.~(\ref{fact}) suppressed by at least $\Lambda_{\rm QCD}^2/p_\perp^2$ at large $p_\perp$. We
emphasize that our recombination mechanism generates the asymmetry by a perturbative mechanism in
which a $\overline{b}$ combines with a light quark participating in the hard scattering process. Thus,
it is suppressed at large $p_\perp$ only by $\Lambda_{\rm QCD} m_b/p_\perp^2$.  Our process is
therefore more important than the beam drag effect for $p_\perp \gg m_b$. Even for $p_\perp \sim m_b$,
our process may be numerically more important at large rapidity, because it allows most of the
momentum of a valence parton to be delivered to the heavy meson.

\section{Heavy-quark recombination cross section}

\begin{figure}[!t]
 \centerline{\epsfysize=7.0truecm \epsfbox{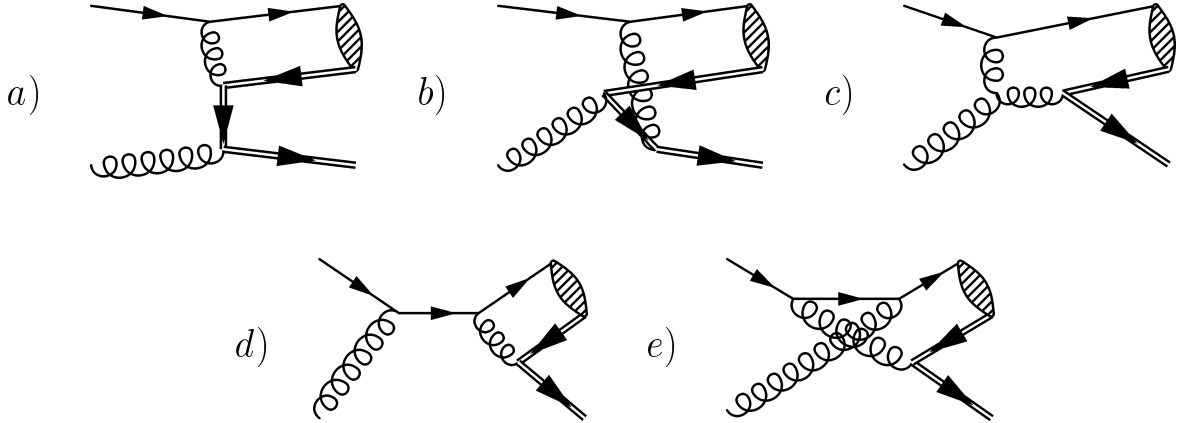}  }
% \centerline{\epsfysize=7.0truecm \epsfbox{basym_fig1.ps}  }
 {\tighten
\caption[1]{Diagrams for $\overline{b}q$ recombination into a $B$ meson.
Light quarks are single lines, heavy quarks are double lines
and the shaded blob is the $B$ meson.}
\label{lo} }
\end{figure}

This paper presents calculations of the $\overline{b}q$ recombination contribution to the
inclusive cross section for $B$ hadrons at the Tevatron. We will focus on the rapidity dependence
of $B$ hadron production. The rapidity dependence of the inclusive cross section for
$\overline{B}$ hadrons can be obtained by replacing $y \rightarrow -y$.  The leading order
diagrams contributing to $q g\rightarrow(\overline{b}q)+b$ are shown in Fig.~\ref{lo}a-e. The
single lines represent light quarks and double lines are heavy quarks. The shaded blob represents
the transition amplitude for the $(\overline{b} q)$ to hadronize into a state including a $B$
hadron.

Let us denote the momentum of the initial-state light quark as $p$,
the initial-state gluon as $l$, the
final-state light quark as $p_q$, the momentum of the $\overline{b}$
(which ends up in the $B$) as
$p_b$, and the momentum of the $b$ quark as $k$. In the rest frame
of the $B$ hadron, $p_b$ has a large
component of $O(m_b)$ while all components of $p_q$ are $O(\Lambda_{\rm QCD})$.
Therefore, in the perturbative
amplitude we can take $p_q\rightarrow 0 $ up to corrections of
$O(\Lambda_{\rm QCD}/m_b)$. The $O(p_q^{-1})$
piece of the  amplitude is:
\bea\label{amp}
{\cal M}_{(\overline{b}q)} &=& g_s^3\, \bar{u} (p_q)
\left( {A_1  \over 2 p \cdot p_q}
+ {A_2  \over 2 l \cdot p_q}\right) v(p_b) \\
  &=&  g_s^3 {\rm Tr} \left[ v(p_b) \bar{u}(p_q)
  \left( {A_1  \over 2 p \cdot p_q }
+ {A_2  \over 2 l \cdot p_q}\right) \right] \nonumber \, ,
\eea
where $A_1$ and $A_2$ are given by:
\bea\label{amp2}
A_1 &=& T^b \gamma^\mu u(p)\cdot\bar{u}(k)
\left( T^a T^b \epsl {1\over -\!\lsl \, + \ksl -m_b} \gamma_\mu
+ T^b T^a \gamma_\mu
{1\over \lsl \, - \psl_b -m_b} \epsl \right. \\
&& \left. + i f^{abc}T^c{(\lsl-\psl)\epsilon_\mu
+ 2 p\cdot \epsilon \,\gamma_\mu -2 \,l_\mu \epsl \over (k+p_b)^2 }\right) \, ,
\nonumber \\
A_2 &=& - T^a T^b {\epsl \lsl \,\gamma^\mu  \over (k+p_b)^2 }
u(p)\cdot\bar{u}(k)T^b\gamma_\mu
\nonumber\, .
\eea
The trace in Eq.~(\ref{amp}) is over both Dirac and color indices.
The polarization vector of the gluon
is $\epsilon_\mu$. Contributions to $A_1$ come from diagrams in
Figs.~\ref{lo}a)-c), while $A_2$ comes
from Fig.~\ref{lo}e). Note that Fig.~\ref{lo}d) does not have any
terms that are singular as
$p_q\rightarrow 0$, so we neglect its contribution relative to the
terms kept in Eq.~(\ref{amp2}).

At this point we need to project the partonic amplitude onto a
matrix element that gives the transition
amplitude to a state with a $B$ hadron. Let us assume for the
moment that the $(\overline{b}q)$ evolves into
the exclusive state consisting of a single $B$ meson,
which is possible only if the $(\overline{b}q)$ are in a
color-singlet state. (Later, we will consider the contribution
of $(\overline{b}q)$ evolving into final states
containing additional soft hadrons.) Then we should replace
the spinors in Eq.~(\ref{amp}) with the valence
light-cone wavefunction of the $B$ meson. The formalism we use
is similar to that used for exclusive $B$
decays and form factors
\cite{Pirjol:2000gn,Korchemsky:2000qb,Grozin:1997pq,Beneke:2001wa,Grinstein:2000pc,Beneke:2000ry}.
We will employ the conventions of \cite{Grozin:1997pq,Beneke:2001wa}.
The light-cone wavefunction of the $B$ meson is:
\bea\label{lc}
\langle B(v)|\bar{q}_{\beta,j}(z) b_{\alpha,i}(0)|0\rangle &=&
{\delta_{ij} \over N_c}
{i f_B m_B\over 4}\left[{1-\vsl\over 2}\left\{ 2 \bar{\phi}_+(t) -
\zsl{\bar{\phi}_-(t) - \bar{\phi}_+(t) \over t}
\right\}\gamma_5\right]_{\alpha \beta} .
\eea
In Eq.~(\ref{lc}), $z$ is a light-like vector, $p_b = m_b v$
and $t=v \cdot z$. The $\alpha,\beta$ are
Dirac indices, $i,j$ are color indices.  For $(\overline{b} q)$
in a $^3S_1^{(1)}$ state,  we should
replace the $B$ meson with a $B^*$ meson and the $\gamma_5$
with $\epsl$, where $\epsilon^\mu$ is the
polarization vector of the $B^*$. The right hand side
is the most general Lorentz decomposition of the
non-local matrix element consistent with heavy quark symmetry.
The functions $\bar{\phi}_\pm(t)$ are
defined so that $\bar{\phi}_\pm(0)=1$ and $f_B$ is the
$B$ meson decay constant. The amplitude for
producing the $B$ meson is obtained by making the
following substitution in the matrix element for the
partonic cross section in Eq.~(\ref{amp}) \cite{Beneke:2001wa}:
\bea
v_i(p_b) \bar{u}_j(p_q) \longrightarrow && \\
-{\delta_{i j} \over N_c} {i f_B\over 4}&&{\psl_b - m_b\over 2}
\int_0^\infty d\omega
\left( 2 \phi_+(\omega) + \int_0^\omega d\eta
\left( \phi_-(\eta)-\phi_+(\eta)\right) \gamma^\mu
	{\partial \over \partial p_q^\mu }\right)
\gamma_5 {\bigg |}_{{\large p_q =\omega \, v}}
\, . \nonumber
\eea
The functions $\phi(\omega)$ are the Fourier transforms of $\bar{\phi}(t)$.
When making this
substitution, $p_q$ should be replaced with $\omega v$
only after taking derivatives on the amplitude. It is
straightforward  to take derivatives and integrate by parts to obtain:
\bea
{\cal M}_B = {g_s^3 \over N_c} {\rm Tr}\left[(\psl_b- m_b)\gamma_5
\left\{ \left( f_+ -
{\psl \over 2 p\cdot v} f_-\right){A_1\over 2 p\cdot v} +
\left(f_+ - {\lsl \over 2 l\cdot v} f_-\right)
{A_2\over 2 l\cdot v} \right\}\right] \, ,
\eea
where the $f_\pm$ are defined by
\bea
f_+ &=& {f_B\over 4}\int_0^\infty d\omega {\phi_+(\omega) \over \omega}  \, ,\\
f_- &=& {f_B\over 4}\left(\int_0^\infty d\omega
{\phi_-(\omega)-\phi_+(\omega)\over \omega}
+ \phi_-(0) \right) \, . \nonumber
\eea
In the last equation, we have used $\phi_+(0) =0$. Since $\phi_-(0)\neq 0$,
the integral in the definition
of $f_-$ is divergent \cite{Grozin:1997pq,Beneke:2001wa}.
This divergence is cut off by the transverse
momentum of the light quark in the $B$
meson. If the cross section depended on $f_-$ it would indicate
that the recombination mechanism is
sensitive to soft physics. Fortunately, this is not the case.
It is possible to show that  $\psl A_1 =\lsl
A_2 =0$. For example,
\bea
\lsl A_2 \sim \lsl \epsl \lsl = 2 \epsilon\cdot l \lsl - \epsl \, l^2 =  0 \, .
\eea
Thus all terms proportional to $f_-$ give a vanishing contribution
to the amplitude.
The final result for the amplitude for the exclusive recombination
production of a $B$ meson is:
\bea\label{finalans}
{\cal M}_B &=&  {g_s^3 m_b f_+ \over N_c}{\rm Tr}
\left[ (\psl_b-m_b)\gamma_5 \left(
{A_1 \over 2 p \cdot p_b } + {A_2 \over 2 l \cdot p_b}\right) \right] \, .
\eea
This amplitude can be obtained from the partonic amplitude of
Eq.~(\ref{amp}) by making a few simple
substitutions. For the Dirac spinors, we make the substitution
\bea\label{sub}
v_i(p_b) \bar{u}_j(p_q) \longrightarrow {\delta_{i j} \over N_c}
m_b f_+ (\psl-m_b)\gamma_5 \, ,
\eea
and then we replace all other occurrences of $p_q$ in Eq.~(\ref{amp})
with $p_b$. Since $f_B$ scales like $\Lambda_{\rm QCD}^{3/2}/m_b^{1/2}$
and $\phi_+(\omega)$ scales like $\Lambda_{\rm QCD}^{-1}$,
we conclude that $f_+$ scales like $(\Lambda_{\rm QCD}/m_b)^{1/2}$.

Note that the final expression obtained for the production matrix
element is very similar to what would be
obtained in a quarkonium production calculation.
If we replaced the light quark with a charm quark in the
above calculation, then the outgoing momentum would be $p_q= r p_B$
and  $p_b=(1-r)p_B$, where
$r=m_c/(m_c+m_b)$ and $p_B$ is the $B_c$ momentum.
The projection operator for the $B_c$ is
\bea
v(p_b)_i \bar{u}_j(p_q) \longrightarrow
\delta_{i j}{f_{B_c} \over 12}(\psl_B -m_{B_c})\gamma_5 \, .
\eea
Thus the matrix element for $B$  production via $(\overline{b}q)$
in a $^1S_0^{(1)}$ state can be obtained by
taking the $r\rightarrow 0$ limit  of the matrix element
for $B_c$ production  and replacing $f_{B_c}/(4
r)\rightarrow f_+ m_b$. The resulting cross sections are
equivalent to modelling the $B$ meson as
a nonrelativistic bound state with a light constituent quark.

In calculating the inclusive cross section, one must also include
processes in which the $(\overline{b} q)$
hadronizes into multiparticle final states, such as $B\pi, B\pi\pi$, etc.
After squaring the matrix element, the exclusive cross
section is proportional to $f_+^2$. The field theoretic
expression for $f_+^* \,f_+$ involves
a projection operator $\mid\!B(v)\rangle \langle B(v)\!\mid$.
The inclusive cross section will
instead involve the inclusive projection operator
\bea\label{incpro}
\sum_X \mid\!B(v)+X\rangle \langle B(v)+X\!\mid \, ,
\eea
where the sum is over all possible soft hadrons $X$ in the
rest frame of the $B$. After manipulations
analogous to the exclusive case, we would ultimately arrive
at an expression proportional to a double moment of some
density matrix instead of $\phi^*(\omega^\prime)\phi(\omega)$.
The bottom line is that the inclusive cross
section for color-singlet $\overline{b}q$ recombination
can be obtained  from the exclusive cross section
simply by replacing $f_+^2$ by a different nonperturbative
parameter $\rho_1$. The parameter $\rho_1$
is proportional to the probability for the color-singlet $(\overline{b}q)$
to evolve into any state containing the $B$
hadron. Like $f_+^2$, the parameter $\rho_1$ scales like
$\Lambda_{\rm QCD}/m_b$.

If the vacuum saturation approximation is applied to the sum over soft hadrons in Eq.~(\ref{incpro}), it
would give $f_+^2$ as an estimate for $\rho_1$. In quarkonium production \cite{Bodwin:1995jh}, the vacuum
saturation approximation can be applied to color-singlet matrix elements with the appropriate angular
momentum quantum numbers.  It is a controlled approximation with corrections that scale as $v^4$. In the
case of $B$ hadrons, the vacuum saturation approximation is not a controlled approximation and it is likely
to drastically underestimate the value of $\rho_1$.  There is no reason  to expect the contributions from $B
\pi$, $B \pi \pi$, etc. to be suppressed relative to $B$. The extraction of the parameter $f_+$ from
exclusive $B$ decays could be used to place a lower bound on $\rho_1$, but we emphasize that it should be
regarded as a weak lower bound.

Our final result for the color-singlet $\overline{b} q$ recombination
contributions to the inclusive cross section for $B$ hadron
production in $q g$ collisions is given by
\bea\label{ss}
{d \hat{\sigma} \over dt}[\overline{b}q({^1S_0^{(1)}})] &=&
{2 \pi^2 \alpha_s^3 \over 243} {m_b^2 \over S^3}\left[-{64\,U \over S}
+{m_b^2\over T}\left(79 -{112 \,U\over T} -{64 \,U^2 \over T^2}\right)\right.\\
&& \left.\qquad \qquad \qquad -{16\,m_b^4 \,S \over U\, T^2}
\left(1- {8\,U\over T} \right)\right] , \nonumber
\\
{d \hat{\sigma} \over dt}[\overline{b}q({^3S_1^{(1)}})]   &=&
{2 \pi^2 \alpha_s^3 \over 243}
{m_b^2 \over S^3}\left[-{64\, U \over S}
\left(1+{2\,U^2\over T^2}\right)\right.  \\
&& \left. \hspace{-1 cm}- {m_b^2\over T}\left(19 -{28\,T\over U} + {368\,U \over T}
+{4\,T^2\over U^2}  -{64\,U^2\over T^2}\right)-{48\,m_b^4 \,S \over U\, T^2}
\left(1 - {8\,U\over T} \right)\right] \, \nonumber,
\eea
where we have defined $S=(p+l)^2$,
$T= t - m_b^2 =(p-p_b)^2-m_b^2 = - 2 p\cdot p_b$ and
$U=(l-p_b)^2-m_b^2 = - 2 l\cdot p_b$.
To understand the behavior of the cross section, it is useful to express the variables $S$, $T$ and $U$ in
terms of $p_\perp$ and $\Delta y = \overline{y}-y$, where $\overline{y}$ and $y$ are the rapidities of
the $\overline{b}$ and $b$, respectively:
\bea
S &=& 2(p_\perp^2+m_b^2)(1+{\rm cosh} \Delta y )\nonumber \, , \\
T &=& -(p_\perp^2+m_b^2)(1+e^{-\Delta y}) \nonumber \, , \\
U &=& -(p_\perp^2+m_b^2)(1+e^{\Delta y}) \nonumber \, .
\eea
At fixed $\Delta y$, $d\sigma/dt = \rho_1 \, d\hat{\sigma}/dt \sim \Lambda_{\rm QCD}
m_b/(p_\perp^2+m_b^2)^3$. For large $p_\perp$, the cross sections $d\sigma/dt$ for recombination scale
like $\Lambda_{\rm QCD} m_b/p_\perp^6$, while the partonic cross sections that appear in the
$\overline{b}$ fragmentation formula in Eq.~(\ref{fact}) scale like $1/p_\perp^4$. Thus, recombination
is suppressed at large $p_\perp$ by $\Lambda_{\rm QCD} m_b/p_\perp^2$ relative to $\overline{b}$
fragmentation, consistent with the QCD factorization theorems. Although the recombination  cross
sections are peaked in the forward region, they are always finite as the $b$ quark mass cuts off all
infrared divergences.

It is also possible for a color-octet $(\overline{b}q)$ to hadronize into a state with a $B$ plus soft
hadrons. To obtain the color-octet production cross section starting from the partonic amplitude in
Eq.~(\ref{amp}), we first make the substitution
\bea\label{sub8}
v_i(p_b) \bar{u}_j(p_q) \rightarrow \sqrt{2\over N_c}  T^a_{i j}\,
m_b f_+^8(\psl-m_b)\gamma_5 \, ,
\eea
and replace all other occurrences of $p_q$ with $p_b$. After squaring the matrix element and summing
over colors in the final state, $(f_+^8)^2$ is replaced with $\rho_8$, a factor that is proportional to
the probability for a $(\overline{b}q)$ in a color-octet state to hadronize into a state including the
$B$ hadron. The normalization in Eq.~(\ref{sub8}) is chosen so that $\rho_8/\rho_1$ is the ratio of the
probabilities for a color-octet $(\overline{b}q)$ and a color-singlet $(\overline{b}q)$ to hadronize
into a $B$ hadron. Since both $\rho_1$ and $\rho_8$ are $O(\Lambda_{\rm QCD}/m_b)$, we might expect
this ratio to be approximately 1. Our results for the color-octet $\overline{b} q$ recombination
contributions to the inclusive cross section for $B$ hadron production in $q g$ collisions are:
\bea\label{os}
{d \hat{\sigma} \over dt}[\overline{b}q({^1S_0^{(8)}})]  &=&
{4 \pi^2 \alpha_s^3 \over 243}{m_b^2 \over S^3}\left[-{U \over S}\left(4 + {9\,T \over U}
+ {9\,T^2 \over U^2} \right) \right. \\
&& \left. -{m_b^2\over 2 T}\left(79 +{14 \,U\over T} +{8 \,U^2 \over T^2}
-{18 T^2 \over U^2} \right)
+{\,m_b^4 \,S \over U\, T^2}\left(8+ {9 T \over U}+{8\,U\over T}
\right)\right] \, , \nonumber
\\
{d \hat{\sigma} \over dt}[\overline{b}q({^3S_1^{(8)}})]  &=&
{4 \pi^2 \alpha_s^3 \over 243} {m_b^2 \over S^3}\left[-{U \over S}
\left(22+{9\, T\over U}+{18 \,U \over T}+{9\,T^2\over U^2}
+{8\,U^2\over T^2}\right) \right.  \\
&&\left.\hspace{-0.25 cm}- {m_b^2\over 2 T}\left(233 +{10\,U\over T}
+ {316\,T \over U} + {266\,T^2\over U^2}-{8\,U^2\over T^2}\right)
+{3\, m_b^4 \,S \over U\, T^2}\left(8+ {9\,T\over U}
+{8\,U \over T}\right) \right] \, \nonumber .
\eea

\begin{figure}[!t]
 \centerline{\epsfysize=16.0truecm \begin{turn}{270} \epsfbox{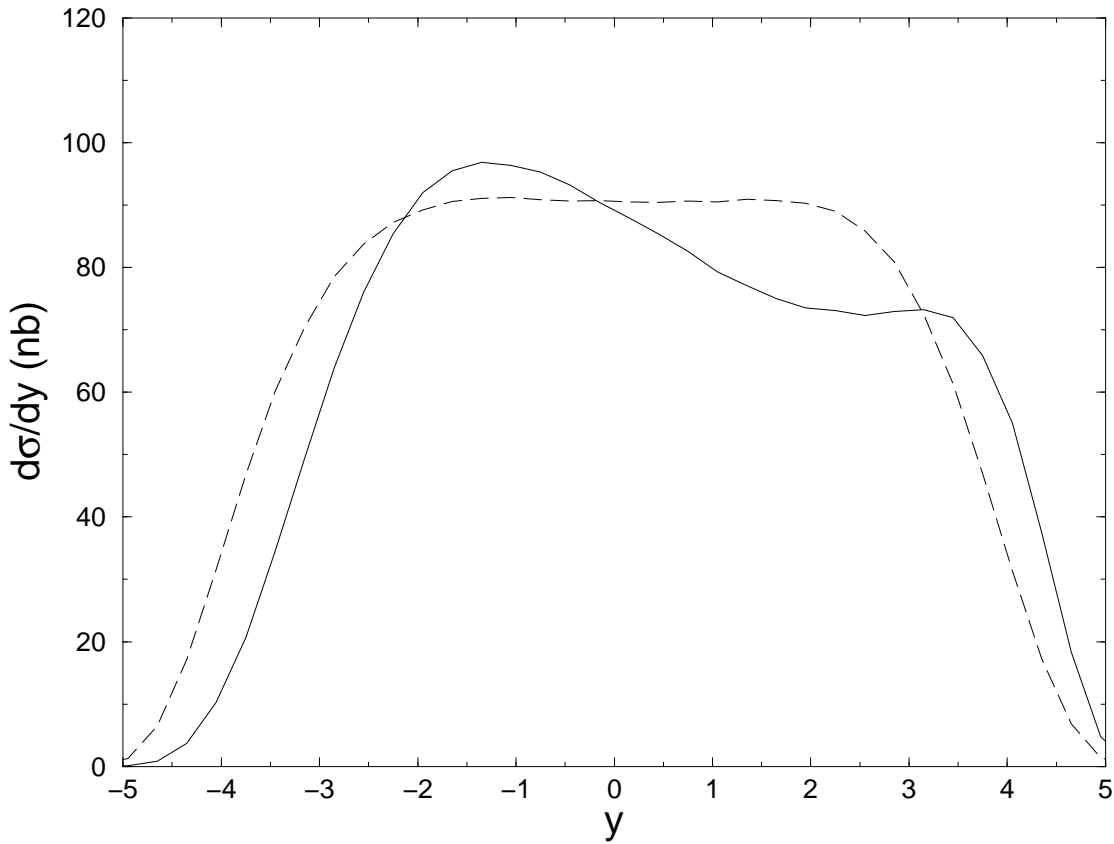}
%  \centerline{\epsfysize=16.0truecm \begin{turn}{270} \epsfbox{basym_fig2.ps}
  \end{turn}  }
  \vspace{0.25 in}
 {\tighten
\caption[1]{
Color-singlet (solid) and color-octet (dashed) contributions
to the rapidity distribution of $B$ hadrons in
$p\overline{p}$ collisions at $\sqrt{s}=1.8$ TeV,
with a $p_\perp > 5\,{\rm GeV}$ cut. We have chosen
$\rho_1=0.1$ and $\rho_8=0.17$,
so color-singlet and color-octet cross sections are roughly equal at central
rapidity. }
\label{singoct}}
\end{figure}

The normalization of the $\overline{b} q$ recombination contributions to the inclusive cross section
for $B$ hadron production depend on the parameters $\rho_1$ and $\rho_8$. Since their values are not
known, we will focus on the qualitative aspects of the rapidity distributions in $p\overline{p}$
collisions.  Note that $\overline{b} q$ recombination can actually contribute to $B$ hadron production
in two ways: \bea\label{bplus} &a)& \qquad \sum_n d\hat{\sigma}[q g \rightarrow (\overline{b} q)^n +b]
\,  \rho_n[(\overline{b}q)^n\rightarrow B ]\, ,\\ &b)& \qquad \sum_{\overline{q},n}
d\hat{\sigma}[\overline{q} g \rightarrow (b\overline{q})^n+ \overline{b}]\, \rho_n \otimes
D_{\overline{b}\rightarrow B} \nonumber \, . \eea In $a)$, the $(\overline{b}q)$ hadronizes into the
$B$ hadron. In $b)$, the $(b\overline{q})$ hadronizes into a $\overline{B}$, and the $B$ comes from the
fragmentation of the associated $\overline{b}$. In $p \bar p$ collisions, both processes in
Eq.~(\ref{bplus}) lead to asymmetric $B$ hadron rapidity distributions. As discussed earlier, process
$a)$ yields $B$ hadrons strongly peaked in the proton direction. The most important contribution to
process $b)$ comes when the initial-state $\overline{q}$ is a  $\overline{u}$ or $\overline{d}$ from
the antiproton, and the resulting $B$ hadron cross section is peaked at negative rapidity. Since $a)$
produces $B$'s with positive rapidity while $b)$ produces $B$'s with negative rapidity, the asymmetry
in the rapidity distribution tends to get washed out when the two effects are combined. The extent to
which this happens  depends on the color state of the $(\overline{b} q)$ and $(b\overline{q})$. This is
illustrated in Fig.~\ref{singoct}, which shows the rapidity distribution of $B$ hadrons from
$\overline{b} q$ recombination in $p\overline{p}$ collisions at $\sqrt{s}=1.8$ TeV, with a $p_\perp >
5\,{\rm GeV}$ cut. We have set $m_b = 4.75$ GeV and used the CTEQ5L \cite{Lai:2000wy} parton
distribution functions. We used the one-loop running formula for $\alpha_s(\mu_R)$ with parameter
$\Lambda^{(5)}= 0.146$ GeV and set the renormalization and factorization scales to $(m_b^2 +
p_\perp^2)^{1/2}$. The color-singlet and color-octet $\overline{b} q$ recombination cross sections are
the solid and dashed lines, respectively. The unknown parameters $\rho_1$ and $\rho_8$ are
$O(\Lambda_{\rm QCD}/m_b)$ and therefore are expected to be less than 1. We have chosen $\rho_1=0.1$
and  $\rho_8=0.17$. Their relative normalization is such that the two distributions are approximately
equal at central rapidity. Color-singlet recombination produces an asymmetric rapidity distribution
with a small peak in the forward direction from process $a)$ and a large peak in the backward direction
from process $b)$. In color-octet production, the processes $a)$ and $b)$ combine to yield a
surprisingly symmetric rapidity distribution. Thus the $B$ rapidity asymmetries can give important
information on the relative importance of color-octet and color-singlet production mechanisms.

\begin{figure}[!t]
\centerline{\epsfysize=16.0truecm
  \begin{turn}{270} \epsfbox{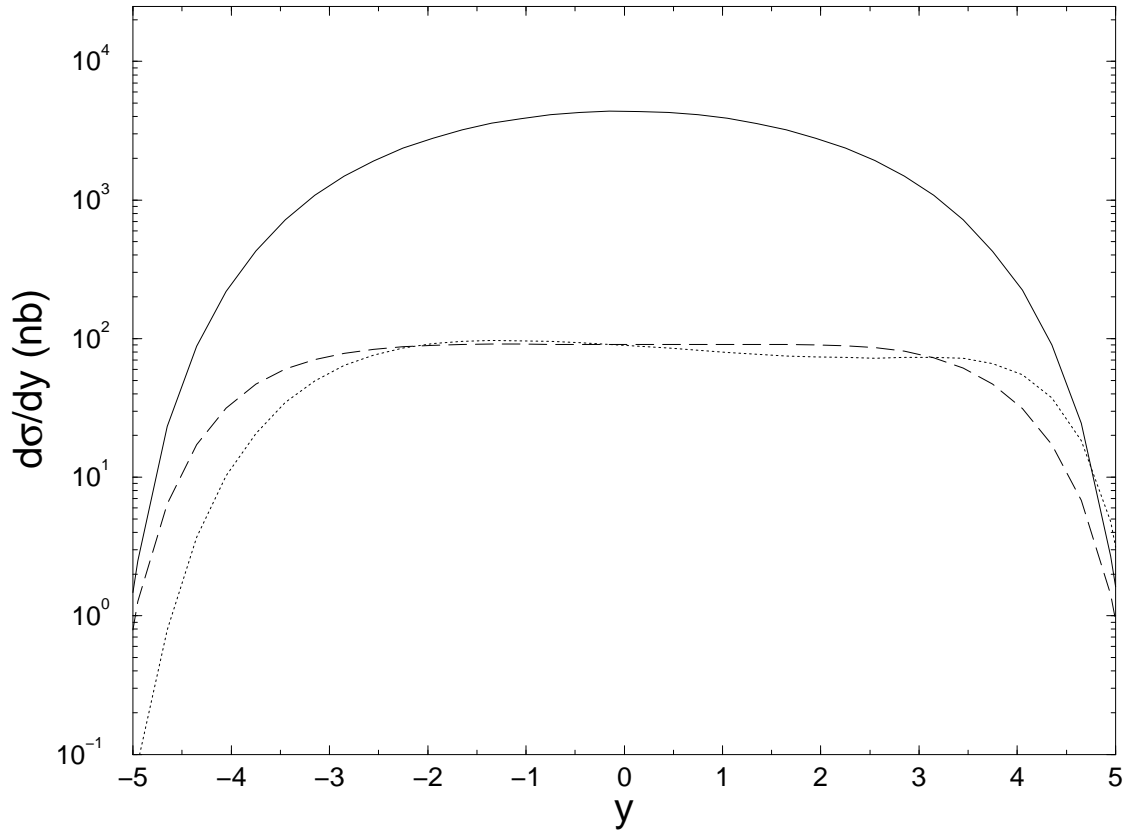}\end{turn}  }
% \begin{turn}{270} \epsfbox{basym_fig3.ps}\end{turn}  }
\vspace{0.25 in}
{\tighten
\caption[1]{Rapidity distribution of $B$ hadrons in $p\overline{p}$
collisions at $\sqrt{s}=1.8$ TeV, with a $p_\perp > 5\,{\rm GeV}$ cut.
Solid line is NLO $b$ fragmentation contribution,
dotted and dashed lines are color-singlet and color-octet contributions.
The parameters $\rho_1$ and $\rho_8$ are the same as in
Fig.~\ref{singoct}. }
\label{dsdy}}
\end{figure}

\section{Comparison with fragmentation mechanism}

Next we compare the $\overline{b} q$ recombination cross sections with the NLO perturbative QCD
calculations of $\bar b$ fragmentation production of $B$ hadrons in $p\overline{p}$ colliders. The
rapidity distributions are compared in Fig.~\ref{dsdy}. The solid line shows the $\bar b$ fragmentation
contribution, which is obtained by using the LO cross section for $\overline{b}$  production convolved
with CTEQ5L parton distribution functions \cite{Lai:2000wy}, then multiplied by a K factor of 2.0 to
give a normalization consistent with the NLO calculation \cite{Beenakker:1991ma}. In
\cite{Beenakker:1991ma,Abe:2000ac}, it is demonstrated that the shapes of the NLO and LO rapidity
distributions are nearly identical, so multiplying the LO cross section by a K factor gives an adequate
representation of the NLO calculation. The partonic cross section should be convolved with a $b$ quark
fragmentation function. We have checked using the Peterson fragmentation function
\cite{Peterson:1983ak} that the shape of the rapidity distribution is essentially unaffected by this
convolution. Also depicted in Fig.~\ref{dsdy} are the $\overline{b} q$ recombination cross sections for
$B$ hadrons, including both processes in Eq.~(\ref{bplus}). The dotted and dashed lines show the
rapidity distributions for color-singlet and color-octet $\overline{b} q$ recombination, respectively.
The parameters $\rho_1$ and $\rho_8$ are the same as in Fig.~\ref{singoct}.  We emphasize that the
normalization of the recombination contributions in Fig.~\ref{singoct} is uncertain, and that only
their shapes are determined.  It is clear that the recombination cross section can be as large as the
fragmentation contribution for sufficiently large rapidity, however the point at which this occurs
depends sensitively on $\rho_1$ and $\rho_8$.

\begin{figure}[!t]
 \centerline{\epsfysize=16.0truecm \begin{turn}{270} \epsfbox{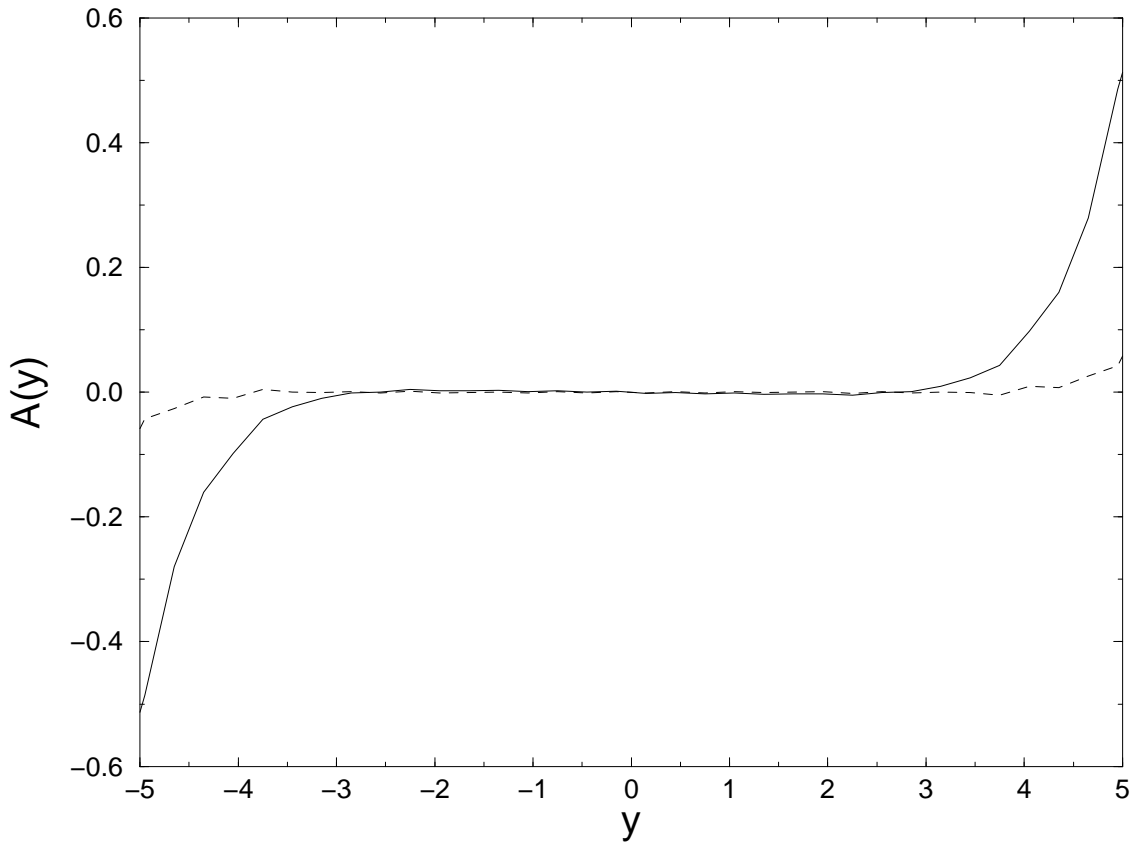}
%  \centerline{\epsfysize=16.0truecm \begin{turn}{270} \epsfbox{basym_fig4.ps}
  \end{turn} }
 \vspace{0.25 in}
 {\tighten
\caption[1]{Rapidity asymmetry for $B$ hadrons in $p\overline{p}$ collisions
at $\sqrt{s}=1.8$ TeV, with a $p_\perp > 5\,{\rm GeV}$ cut.  The solid line is color-singlet,
dashed line is color-octet. The parameters $\rho_1$ and $\rho_8$ are the same as in
Fig.~\ref{singoct}. }
\label{asym}}
\end{figure}

It is also interesting to look at the rapidity asymmetry of $B$ production,
which
is defined by
\bea
A(y) = {\sigma_{B}(y) -\sigma_{B}(-y)\over \sigma_{B}(y) +\sigma_{B}(-y)} \, .
\eea
This is shown in Fig.~\ref{asym}. Color-singlet recombination, shown by the solid line, only gives a
noticeable asymmetry for $|y| > 3$, while color-octet recombination, shown by the dashed line, gives a
negligible asymmetry for $|y| < 5$. For the values of $\rho_1$ and $\rho_8$ chosen in this paper, the
asymmetry is unobservable at the rapidities currently probed by the $\D0$ experiment. The asymmetries
can of course be larger or smaller depending on the values of $\rho_1$ and $\rho_8$.

In order to assess how important the $\overline{b} q$ recombination mechanism is in the rapidity
regions that are accessible experimentally,  one needs to know the normalization of the cross
section, which is determined by the parameters $\rho_1$ and $\rho_8$. Unfortunately, these
parameters are unknown. The vacuum saturation approximation gives the weak lower bound $\rho_1 >
f_+^2$. If one makes the mild assumption that the light-cone wavefunction $\phi_+(\omega)$ is
positive definite, it is possible to place a lower bound : $f_+ \geq 0.1$
\cite{Korchemsky:2000qb}.\footnote{We thank Dan Pirjol for discussions on this point.}  The
light-cone wavefunction models employed in  \cite{Korchemsky:2000qb} give estimates for $f_+$ in the
range $0.1-0.2$. We conclude that $\rho_1 > 0.01$.

The actual determination of $\rho_1$ and $\rho_8$ would require phenomenological analysis of $B$
hadron production in other high energy physics processes. We can extract estimates of these
parameters from a constituent quark model for heavy-meson production developed by Berezhnoy, Kiselev
and Likhoded \cite{Berezhnoi:1997st,Berezhnoy:1999rd,Berezhnoy:2000yj,Berezhnoy:2000ji}. In this
model, the $B$ hadron cross section is obtained from a factorization formula analogous to
Eq.~(\ref{onia}) for $\overline{b}c$ hadrons by replacing  $m_c$ with a constituent quark mass,
$m_q$, for the light quark and computing the parton cross section to leading order in perturbative
QCD. In the simplest form of this model, only the color-singlet S-wave matrix element, $\langle
{\cal O}^B_1\rangle$, is kept and it is estimated from quark model wave-functions
\cite{Berezhnoi:1997st}. This model was previously used to calculate heavy quark fragmentation
functions \cite{Braaten:1995bz}. However, the model can easily be extended by relaxing the vacuum
saturation approximation for $\langle {\cal O}^B_1\rangle$ and including a color-octet S-wave matrix
element $\langle {\cal O}^B_8\rangle$ \cite{Berezhnoy:1999rd,Berezhnoy:2000yj,Berezhnoy:2000ji}. The
parton cross sections in this model can include fragmentation, recombination, and other
contributions, depending on the high energy process being considered. All the contributions are
calculated in terms of the two matrix elements $\langle {\cal O}^B_1\rangle$ and $\langle {\cal
O}^B_8\rangle$. Our recombination cross sections can be obtained from calculations using this model
by isolating the recombination contributions and then keeping the leading terms in the limit $r
=m_q/(m_q+m_b) \rightarrow 0$. We can therefore use the matrix elements of
\cite{Berezhnoy:1999rd,Berezhnoy:2000yj,Berezhnoy:2000ji} to estimate our parameters $\rho_1$ and
$\rho_8$.

The simplest way to determine the values of $\rho_1$ and $\rho_8$ that correspond to the matrix
elements $\langle {\cal O}^B_1\rangle$ and $\langle {\cal O}^B_8\rangle$ of
\cite{Berezhnoy:1999rd,Berezhnoy:2000yj,Berezhnoy:2000ji} is to compare their calculations of $D^*$
fragmentation functions \cite{Berezhnoy:1999rd} with what we would get if we applied our formalism.
For the parameters $\rho_n[c \bar q \to D^*]$, we find $\rho^{D^*}_{1,8} = 3
\langle {\cal O}_{1,8} \rangle /(4 m_c m_q^2)$, where $m_q = 0.3$ GeV is a constituent quark mass.
>From  studies of photoproduction and electroproduction \cite{Berezhnoy:2000yj,Berezhnoy:2000ji}, they
extract $\langle {\cal O}_1 \rangle + \langle {\cal O}_8 \rangle/8$ = 0.25 GeV$^3$ and $\langle {\cal
O}_8 \rangle$ = 0.33--0.49  GeV$^3$. The corresponding values of the parameters in our formalism are
$\rho^{D^*}_1 + \rho^{D^*}_8/8 = 1.4, \rho^{D^*}_8 = 1.8 - 2.8$. We can estimate the corresponding
parameters $\rho_{1,8}^{B^*}$ for $B^*$ production by multiplying by a factor of $m_c/m_b$. This is
because $\rho^{B*}_{1,8}$ scale with the heavy quark mass as $1/m_b$. Note that we are assuming that
the moment of the  light-cone wavefunction appearing in the definition of $f_+$ does not exhibit a
strong dependence on the heavy quark mass between the scale $m_c$ and $m_b$. The resulting estimates
are $\rho_1 \approx 0.4$ and $\rho_8 \approx 0.8$, which are larger than the values we used in the
figures.

So far we have focused exclusively on the contributions in which the parton that combines  with the
$\overline{b}$ is the light quark $q$. There are also processes in which a $\overline{q}$ participates
in recombination:
\bea\label{other}
d\sigma[\overline{q} g \rightarrow B+X] &=&
\sum_n d\hat{\sigma}[\overline{q} g \rightarrow (\overline{b}
\overline{q})^n + b]  \, \rho_n[\overline{b} \overline{q} \rightarrow B] \,  .
\eea
The LO diagrams for the process in Eq.~(\ref{other}) are identical to the diagrams shown in
Fig.~\ref{lo}, except the light quark lines are replaced with light antiquark lines. Since the
diagrams for $\overline{b} \overline{q}$ recombination are almost identical to those for
$\overline{b}q$ recombination, we expect that the cross section will be peaked in the direction of the
incoming $\bar q$.  The $B$ hadron will emerge preferentially in the $\overline{p}$ direction rather
than the $p$ direction, so one would expect that the asymmetry would be the reverse of that seen in
Fig.~\ref{asym}. In the case of color-octet production, we saw that the asymmetry was almost
completely washed out by the process $b)$ in Eq.~(\ref{bplus}). Thus, it is difficult to tell how big
the asymmetries from $\overline{b}\overline{q}$ recombination will be without an explicit calculation.
The $\overline{b} \overline{q}$ recombination process will introduce additional nonperturbative
parameters. To lowest order in $\Lambda_{\rm QCD}/m_b$ and $\Lambda_{\rm QCD}/p_\perp$, the
$\overline{b}$ and $\overline{q}$ are in an $L=0$ state. Their color state can be
either the $\overline{6}$ or the $3$ of SU(3). Heavy quark spin symmetry relates the spin
states so there will be two parameters, $\rho_3$ and $\rho_6$, that are analogous to the $\rho_1$ and
$\rho_8$ parameters for $\overline{b} q$ recombination. Intuitively, one might expect
$(\overline{b}q)$ to prefer to hadronize to $B$ mesons and the diquark $(\overline{b}\overline{q})$ to
prefer to hadronize to the $\Lambda_b$ by picking up another light antiquark. This intuition leads one
to expect that $\overline{b} q$ recombination is more important for heavy meson production and that
$\overline{b} \overline{q}$ recombination is more important for heavy baryon production. This would
probably lead to rapidity asymmetries that have opposite signs for $\Lambda_b$ baryons and $B$ mesons.

One could also consider processes in which a gluon from the hard scattering combines with the
$\overline{b}$. These do not produce any interesting asymmetries because the gluon distributions in the
$p$ and $\overline{p}$ are identical. The $(\overline{b}g)$ can be in a $\overline{3}$, $6$ or
$\overline{15}$ of $SU(3)$, so three nonperturbative parameters are required to describe this
recombination mechanism. At large rapidity, the parton cross sections for $(\overline{b}g)$ may be
larger than those for $(\overline{b}q)$. For reasonable values of $\rho_1$ and $\rho_8$, the
$\overline{b} q$ recombination cross section is too small to explain the excess $b$ production at large
rapidity observed by the $\D0$ collaboration \cite{Abbott:2000wu}. However, it is possible that the
discrepancy could be explained by $\overline{b}g$ recombination.

The excess $b$ production at large rapidity observed by the $\D0$ collaboration cannot be explained by
the conventional fragmentation mechanism. We have proposed a perturbative recombination mechanism that
is suppressed by $\Lambda_{\rm QCD} m_b/p_\perp^2$ at central rapidity, but may give an important
contribution to the cross section at large rapidity. In $p\overline{p}$ collisions, our mechanism
generates asymmetries in the rapidity distributions of $B$ hadrons. The observation of rapidity
asymmetries at the Tevatron that are too  large to be explained by string fragmentation models of
hadronization would provide qualitative evidence  for our mechanism. Our mechanism involves several
nonperturbative parameters, so quantitative evidence would require comparing its predictions with data
on $B$ production in other high energy physics experiments. We have calculated the cross sections for
$\overline{b} q$ recombination in $q g$ collisions and studied their effects on $B$ production at the
Tevatron. Future work should involve a complete calculation of the other recombination contributions
discussed in this paper, and an exploration of the implications of heavy-quark recombination
mechanisms for $b$ and $c$ hadron production in $e p$ collisions and fixed target experiments.

We thank S. Fleming, D. Pirjol and D. Soper for useful discussions. This work was supported by the
National Science Foundation under Grant No.\ PHY--9800964 and by the Department of Energy under grant
DE-FG02-91-ER4069.

\end{document}